\documentclass[twocolumn,showpacs,preprintnumbers,amsmath,amssymb,superscriptaddress,10pt,aps,prb]{revtex4-2}
\usepackage{epsfig,amsopn}
\usepackage{graphicx}
\usepackage{epstopdf}
\usepackage{amsmath,amssymb}
\usepackage{amsthm}
\usepackage{enumerate}
\usepackage{xcolor}
\usepackage{fixmath}
\newcommand\mb{\mathbold}

\begin{document}
 \title{Dipolar optical plasmon in thin-film Weyl semimetals}

\author{Debasmita Giri}
\affiliation{Department of Physics, Indian Institute of Technology Kanpur, Kanpur 208016, India}
\author{Dibya Kanti Mukherjee}
\affiliation{Laboratoire  de  Physique  des  Solides,  CNRS  UMR  8502, Universit\`{e} Paris-Saclay,  91405  Orsay  Cedex,  France}
\author{Sonu Verma}
\affiliation{Department of Physics, Indian Institute of Technology Kanpur, Kanpur 208016, India}
\author{H.A. Fertig}
\affiliation{Department of Physics, Indiana University, Bloomington, IN 47405}
\affiliation{Quantum Science and Engineering Center, Indiana University, Bloomington, IN, 47408}
\author{Arijit Kundu}
\affiliation{Department of Physics, Indian Institute of Technology Kanpur, Kanpur 208016, India}
	
\begin{abstract}
In a slab geometry with large surface-to-bulk ratio, topological surface states such as Fermi arcs for Weyl or Dirac semimetals may dominate their low-energy properties. We investigate the collective charge oscillations in such systems, finding striking differences between Weyl and conventional electronic systems. Our results, obtained analytically and verified numerically, predict that the Weyl semimetal thin-film host a single $\omega\propto \sqrt{q}$ plasmon mode, that results from collective, anti-symmetric charge oscillations of between the two surfaces, in stark contrast to conventional 2D bi-layers as well as Dirac semimetals with Fermi arcs, which support anti-symmetric acoustic modes along with a symmetric optical mode. These modes lie in the gap of the particle-hole continuum and are thus spectroscopically observable and potentially useful in plasmonic applications.

\end{abstract}
	
\maketitle

\section{Introduction}
Weyl semimetals (WSM's) are three dimensional topological systems characterized by an even number of band-touching points (Weyl nodes), such that, in the vicinity of these points, the electronic states obey Weyl equations, and as a result are chiral~\cite{WSMreview}.  The unique topology of these systems follows from the fact that the Weyl nodes act as sources or sinks of Berry flux. A remarkable consequence of this becomes apparent in slab geometries of these materials, with surfaces oriented so that the projections of different Weyl points onto the surfaces do not lie upon one another.  In these cases one finds topological ``Fermi arcs'' (FA's), in which the two-dimensional Fermi surface of the slab is fractured into disjoint pieces that reside on different surfaces.  Each arc connects a pair of Weyl nodes of opposite chirality. The states of these Fermi arcs inherit the chirality of the bulk nodes, with velocities that disperse in a quasi-one-dimensional manner.  Examples of such materials include TaAs~\cite{taas}, NbAs~\cite{nbas} and, more recently, Co$_3$Sn$_2$S$_2$, for which Fermi arc modes have been identified in ARPES and quasiparticle interference experiments~\cite{CoSnS1,CoSnS2}.

Closely related to WSM's are Dirac semimetals (DSM's). The electronic structures of these systems host Dirac nodes, which may be understood as a limit in which two Weyl nodes of opposite chirality come together at the same momentum point. Fermi arc states can also be present in a Dirac semimetal slab with Dirac node pairs separated in the two-dimensional momentum space of the slab. In such materials a surface hosts an even number of gapless modes that carry current in opposite directions, with backscattering prohibited when a symmetry protecting them is not violated. 
A possible example of such system are the Cd$_2$As$_{3}$ family of materials~\cite{cdas0}, which support remarkable transport properties~\cite{cdas1,cdas2,cdas3}.

When interactions among electrons are considered, these materials should typically host collective modes, including bulk~\cite{weylpl1,weylpl2} as well as surface plasmons mediated by the Fermi arcs~\cite{weylplsurf1,weylplsurf2,weylplsurf3,weylplsurf4,weylplsurf5,weylplsurf6,weylplsurf7,weylplsurf8}. In contrast to thick systems, where electrons on different surfaces have negligible influence on one another, geometries of these systems with large surface-to-volume ratios, specifically slabs, offer a platform in which the surface states are influential and induce novel properties.
For a thin-film geometry, which is the primary focus of our work, FA states of opposite surfaces can no longer be treated individually and the low-energy Fermi surface, in the two dimension, interpolates states predominantly supported by the two surfaces and the bulk~\cite{rkky}. This intertwining of surface and bulk states raises questions on the nature of the collective modes that these materials can support, how they differ between Dirac and the Weyl semimetals, and how both differ from analogous conventional conducting systems.  A natural paradigm for the last of these is a doped bilayer semiconductor, as might be realized in some heterostructures or double quantum wells.  These systems have been known for some time to support two collective modes analogous to plasmons \cite{twolr1,twolr2,twolr3,twolr4,twolr5,twolr6,twolr7}.  At long wavelengths, one of these corresponds to charge oscillates in the two layers which are in-phase, and disperses as $\sqrt{q}$ (with $q$ the momentum of the excitation).   The other involves antisymmetric charge oscillations, and disperses linearly in $q$, i.e., acoustically.  The non-analytic behavior of the symmetric plasmon mode dispersion is a direct consequence of the long-range nature of the Coulomb interaction.  The acoustic nature of the second mode arises because the long-range part of the interactions is screened by the out-of-phase nature of the density oscillations.

As we show below, plasmons in WSM and DSM slabs have some properties in common with the bilayer semiconductor paradigm, but also display some remarkable differences.  Most dramatically, we find that a WSM slab hosts a {\it single} low-energy plasmon mode dispersing as $\sqrt{q}$, but that, at long wavelengths, the density oscillations are {\it antisymmetric} in across surfaces as is the case for the {\it acoustic} mode of a doped bilayer semiconductor.  This behavior turns out to be a consequence of the opposite chiralities of single-particle modes on the two surfaces, and so is a direct consequence of the unusual topological nature of the Weyl semimetal.  Our prediction offers a new avenue for demonstrating this chirality beyond direct surface transport measurements \cite{Hosur_2013,ShuoWang_2017,Wang_2018}.

In recent years, detection of plasmons and their dispersions in two dimensional systems have become possible using scanning near-field optical microscopy \cite{Chen_2012,Fei_2012}.  Such techniques use nanoprobes to produce and detect the electric field of plasmons, and deduce the plasmon dispersion by observing the wavelength of interference patterns as a function of frequency. These techniques could in principle be applied to thin-film geometries of WSM's and DSM's, and in the former case would only be visible for frequencies above the scale at which the charge antisymmetry becomes sufficiently imperfect that electric fields can escape through the film surfaces and couple to an external probe.  For lower frequencies, the electric fields would be confined within the thin film, making the system a natural waveguide.  This suggests energy transport by the system may be particularly efficient in this low-frequency range.




\section{Heuristic Explanation}
Before presenting results of our detailed analysis, we explain qualitatively how the phenomena described above can emerge in WSM and DSM thin films.
Consider a system with conducting states on opposite surfaces of a slab separated by a dielectric bulk, which we assume in this model to have no qualitative effect on the collective modes.  The resulting system is similar to a pair of interacting two dimensional systems, which, as described above, typically supports a symmetric plasmon ($\sim \sqrt{q}$) mode and an antisymmetric acoustic ($\sim q$) mode \cite{twolr1,twolr2,twolr3,twolr4}.  In the cases of WSM's and DM's these surface states may be modeled as a collection of helical states dispersing linearly in the $\hat{x}$-direction,
\begin{align}
E_s^{(\pm)}(\mb{k})=(\pm) s \hbar v_F  k_x; ~~-k_0 < k_y < k_0.
\label{FAdispersion}
\end{align}
Here $s$ is a (pseudo-)spin index, with $s=1$ on one surface and $s=-1$ on the other for Weyl FA's, and $v_F$ is the Fermi velocity.  The index $s$ may be $\pm 1$ on each surface for Dirac FAs, and the overall $\pm$ sign in Eq. \ref{FAdispersion} applies only to the Dirac case and indicates which surface the arcs lie upon.  For both the Weyl and Dirac cases the FA's are taken for simplicity to lie on straight lines between momenta $k_y = \pm k_0$.

At long-wavelengths the collective modes may be well-described  in the random phase approximation (RPA).  In the case of plasmons these are self-sustained oscillations in which the electron densities respond in the same fashion as non-interacting electrons to an effective potential, generated by the Coulomb interaction, which is induced by the density oscillation.
Writing these response functions as $\chi_1(\mb{q},\omega)$, $\chi_2(\mb{q},\omega)$ for the top and the bottom surfaces, respectively, where $\mb{q} = (q_x,q_y)$ is the surface momentum, the condition for a self-sustaining mode becomes (see Appendix)
 \begin{align}\label{eq:slabcond}
& 1 - V(\mb{q})\big(\chi_1(\mb{q},\omega)
 + \chi_2(\mb{q},\omega) \big)\nonumber\\ &+V(\mb{q})^2(1-e^{-2qL}) \chi_1(\mb{q},\omega) \chi_2(\mb{q},\omega)=0,
 \end{align}
where $L$ is the width of the slab and $V(\mb{q}) = \alpha_c/ \mb{q}$ where $\alpha_c = 2\pi e^2/\epsilon$.

Because of their simple structure, response functions for the FAs of a DSM and a WSM may be written down straightforwardly, which take the form (see Appendic)
  \begin{align}
  \chi_{1,2}^{\rm DSM}&=\frac{ 2k_0 v_F}{2\pi^2\hbar} \frac{q_x^2}{\omega^2 - v_F^2 q_x^2},~  \chi_{1,2}^{\rm WSM}= \frac{\pm k_0}{2\pi^2\hbar} \frac{q_x}{\omega \mp  v_F q_x}.\nonumber
  \end{align}
Note in the limit of small $q$, $\chi^{WSM}_1=-\chi^{WSM}_2$, a direct reflection of the opposite chiralities of the two surface modes.  This property plays an important role in the WSM collective modes. With these expressions, for the DSM case Eq.~(\ref{eq:slabcond}), at small $v_Fq/\omega$,  results in two plasmon modes with dispersions
\begin{align}
\omega^{(1)}_{\rm D} = v_F\sqrt{1+2\alpha_c\beta L}\cos\theta q, ~ \omega^{(2)}_{\rm D} = v_F\sqrt{4\alpha_c\beta}\cos\theta \sqrt{q},\label{eq:plDirac}
\end{align}
where $\theta = \cos^{-1}\frac{q_x}{q}$ and $\beta = \frac{k_0}{2\pi^2\hbar v_F}$. The strong anisotropies in these expressions reflect those of the DSM Fermi arcs, but beyond this are similar to two-dimensional semiconductor bilayers in hosting a symmetric $\sqrt{q}$ mode and an antisymmetric acoustic mode.
By contrast, for the WSM one obtains a single plasmon mode with dispersion
\begin{align}
\omega_{\rm W} = v_F\sqrt{2\alpha_c\beta}\sqrt{1 + \alpha_c\beta L}\cos\theta \sqrt{q}.
\end{align}
Remarkably, the density oscillations on the two surfaces turns out to be {\it antisymmetric} across the surfaces.  The effect is a direct result of the single-particle surface mode chiralities, and in this way reflects the unusual topology of the WSM system (see Appendix). The result is in stark contrast to what is found in the DSM and in conventional semiconductor bilayers.  Because of the antisymmetry, electric fields associated with the WSM plasmon mode will tend to be confined within the interior of the WSM slab.  This suggests that radiative losses by such plasmons will be limited, so that energy transport by them through the slab will be long-lived relative to comparable DSM's and bilayer semiconductor systems.

The simple heuristic model presented here leaves out a number of properties that are relevant to  more realistic models of these systems.  In particular, bulk states, which host a particle-hole continuum of excitations, may dampen the plasmon modes.  This may occur through interactions between the surface and bulk electrons, as well as through their direct coupling at the single-particle state level.  Moreover, the surface states may themselves hybridize for a thin enough slab.  
By a numerical analysis of a more detailed model, we now show that
these modes indeed persist in spite of these effects.

\begin{figure}[t]
\centering
	\includegraphics[width=.44\textwidth]{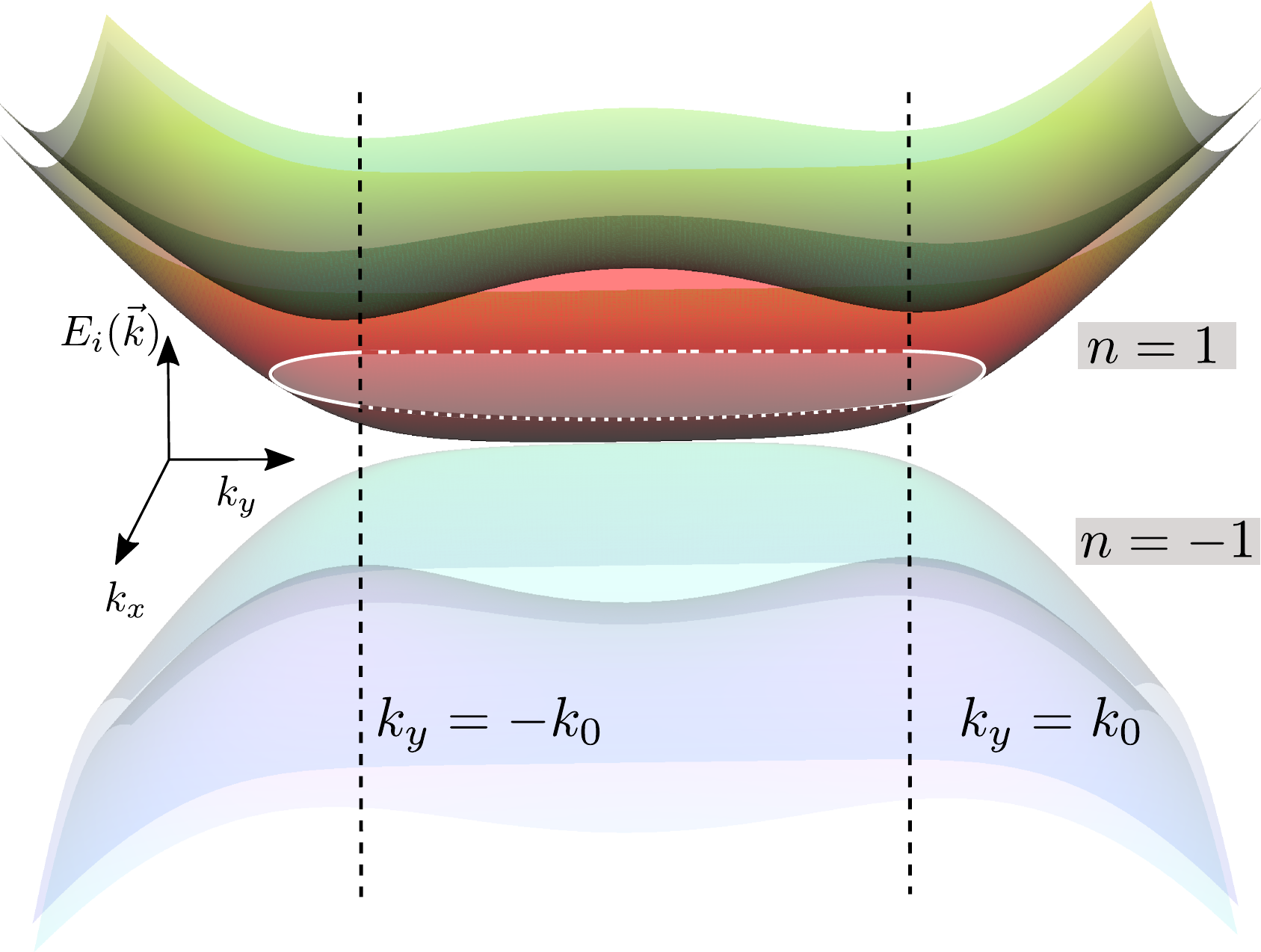}
	\caption{Low-energy bands of the WSM slab. 
 The $n=\pm1$ bands contain all the FA as well as bulk states.
 A Fermi surface at low energy is marked with states at the dotted side is supported by one surface and the states at the dashed-dotted side is supported by the other surface, whereas the states at the solid sides have support dominantly in the bulk.}\label{fig:bands}
\end{figure}

\section{Collective Modes in a Tight-Binding Model}
Our quantitative analysis employs a multiband band model of a semimetal with Weyl points which is block-diagonal in 2$\times$2 blocks, each of which contains a pair of Weyl nodes. The model generalizes to $n$-pairs of Weyl nodes, for which the Hamiltonian consists of $n$-blocks of two band systems. The basic Hamiltonian block for the semimetal may be written as
\begin{align}
 	H_{\eta} &= (\sigma_y q_x - \sigma_x q_z) + \sigma_z M_{\eta}(k_y),\label{eq:Heta}
\end{align}
where $\eta$ maybe $+1$ or $-1$.  In Eq.~(\ref{eq:Heta}) $\mb{\sigma}$ are Pauli-matrices acting on spin amplitudes, and the mass is given by $M_\eta(k_y)=\eta~(m-\cos(k_y))$. We have taken the lattice-spacing $a$ to be unity and have scaled the Hamiltonian by $\hbar v_F /a$, with $v_F$ being the Fermi velocity near the nodes. The momenta are scaled by $1/a$, making all the variables unitless. The value of $m$ determines whether the spectrum is gapped: for $0<m<1$ it contains two Weyl nodes and, for either choice of $\eta$, these Weyl nodes situate at $ \mb{k}=(0, \pm k_0 ,0) $, where $2k_0 = 2\cos^{-1}(m)$ is the momentum separation between them. For a given $\eta$ (say, $\eta=1$), the Hamiltonian Eq.~(\ref{eq:Heta}) breaks time-reversal symmetry. This serves as the simplest model of a WSM. If one retains one block each of the form $H_{\eta=+}$ and $H_{\eta=-}$, which are time-reversal partners, together they serve as a four band model for a DSM.

For a slab geometry with a finite thickness $L$ along the $z$ direction, the electronic states from these Hamiltonians can be obtained by imposing appropriate boundary conditions on the surfaces~\cite{rkky} (see details in the Appendix).  The resulting energy bands are indexed by $n=\pm1, \pm2 \cdots$ ($\pm$ for positive and negative energy bands), for each $\eta$ sector.
We focus upon the case when the chemical potential is positive and the system near charge neutrality, in which case, for the low-energy collective excitations, one may neglect bands other than $n=\pm1$. For such a choice of Fermi energy, the Fermi surface contains the FA states of both surfaces as well as bulk states.

To proceed we expand the second-quantized field operators in eigenstates of the Hamiltonian, $\phi_{\eta,n,\mb{k}}(z)$, as
\begin{align}
&\Psi(\mb{r},z)=\frac{1}{\sqrt{L_x L_y}}\sum_{\mb{k}} \exp(i \mb{k} \cdot \mb{r}) \sum_{\eta,n} \phi_{\eta,n,\mb{k}} (z) c_{\eta,n,\mb{k}}\nonumber
\end{align}
where $c_{\eta,n,\textbf{k}}$ annihilates an electron from $n^{\rm th}$ band in state $\eta,\textbf{k}$ with $\eta=\pm 1$  ($\eta=1$) for a DSM  (WSM).
To find the plasmon modes we consider the density response function (see Appendix)
\begin{align}
\chi(\mb{q},z,z', \omega)=
-\textit i  \int dt d\mb{r}e^{i\mb{q} \cdot \mb{r} +i\omega t}
\langle  \big[\rho(\mb{r},z,t), \rho(0,z',0)] \big\rangle,\nonumber
\end{align}
with $\rho(\mb{r},z,t)= \Psi^{\dagger}(\mb{r},z)\Psi(\mb{r},z)$, the (time-dependent) density operator in the Heisenberg picture.
The poles of $\chi(\mb{q},z,z',\omega)$ denote the values of $\omega$ and $\mb{q}$ where there are collective excitations.  In the time-dependent
Hartree approximation, this response function obeys the equation
\begin{align}\label{eq:finalchi}
\chi(\mb{q},z,z^{\prime},\omega)=& \chi_0(\mb{q},z,z^{\prime},\omega)\nonumber\\
+& \int_{0}^{L}  dz_2  B(\mb{q},z,z_2,\omega) \chi(\mb{q},z_2,z^{\prime},\omega) ,
\end{align}
where
\begin{align}
B(\mb{q},z,z_2,\omega) = \int_{0}^{L} dz_1  V_{\mb{q}}(| z_1-z_2|) \chi_0(\mb{q},z,z_1,\omega)
\end{align}
contains the Coulomb interaction $V_{\mb{q}}(|z|) = \frac{\alpha e^{-q|z|}}{q}$, written in terms of $\alpha$ which we now set to 1~\cite{alpha}.  In these expressions
$\chi_0$ is the non-interacting response function.
To solve these equations,
the integral of the coordinate $z_1$ is performed analytically, while the integral over the coordinate $z_2$ is approximated by a discrete sum, with $\Delta z$ the interval between grid points. The poles of the response function can then be found by solving $\text{det}[ I-\Delta z B(\mb{q},\omega) ]=0$, where $B(\mb{q},\omega)$ is a matrix whose components are given by $B(\mb{q},\omega,z,z')$.

\begin{figure}
\centering
	\includegraphics[width=.49\textwidth]{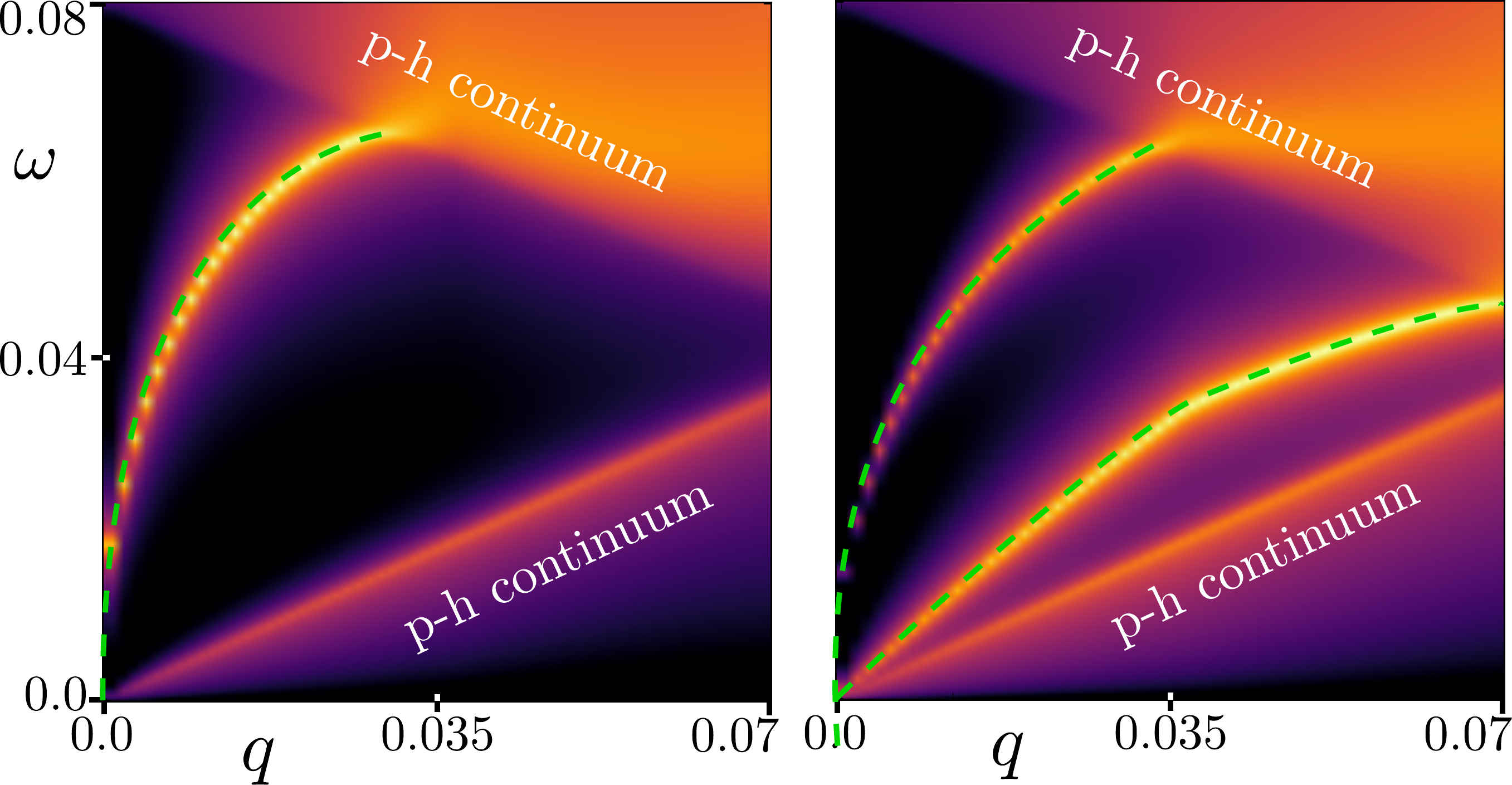}
	\caption{ Particle-hole continuum (lighter area) and the sharp plasmonic modes (marked) as a function of $q$ along the direction of $\tan^{-1}(q_y/q_x) = 60^{{\rm 0}}$. The left and the right plots are for the WSM and DSM thin films, respectively. For numerical results we choose $L=25, \mu=0.04, m=0.5$ and we use 51 divisions along $z$ in finding the $B$ matrix (see Appendix). }
\label{fig:pldisp}
\end{figure}

\section{Numerical results}
Fig.~\ref{fig:pldisp} illustrates typical results from our numerical model.  At low frequencies and wavevectors, sharp modes are visible which are consistent with expectations from the heuristic model discussed above.  Specifically, for the WSM, a single plasmon mode dispersing as $\sqrt{q}$ is apparent, whereas for the DSM, there is in addition an acoustic mode.
An important consideration in obtaining these modes is whether the density response associated with them is truly sharp, as required for a self-sustaining mode.  This can only occur if the particle-hole excitations associated with poles of $\chi_0$, the non-interacting response, are absent for the values of $\mb{q}$ and $\omega$ at which the plasmon modes are present.  It is here that the  bulk states, absent in our heuristic model, have an impact.

The continuum of non-interacting particle-hole excitations in this system consists of two contributions: inter-band and intra-band processes. Intra-band particle-hole excitations exist below any frequency $\omega = v_Fq$, where $v_F$ is the Fermi velocity.  Inter-band excitations have a gap of 2$\mu$ at $q=0$, where $\mu$ is the chemical potential, which drops as $q$ increases.
It is apparent in Fig.~\ref{fig:pldisp} that they leave open a window of wavevectors and frequencies where the plasmon modes enter and remain sharp.  Two comments are worth noting about these particle-hole excitations: (i) they involve {\it both} the non-interacting FA states and the bulk states of the system, and (ii) the relevant particle-hole excitations involve only the bands closest to zero energy; higher energy bands are also present, but only contribute further particle-hole excitations that leave open the region where the plasmon modes are sharp.  We do not include these explicitly in our calculation, as they have no qualitative impact on the results.



\begin{figure}
\centering
	\includegraphics[width=.44\textwidth]{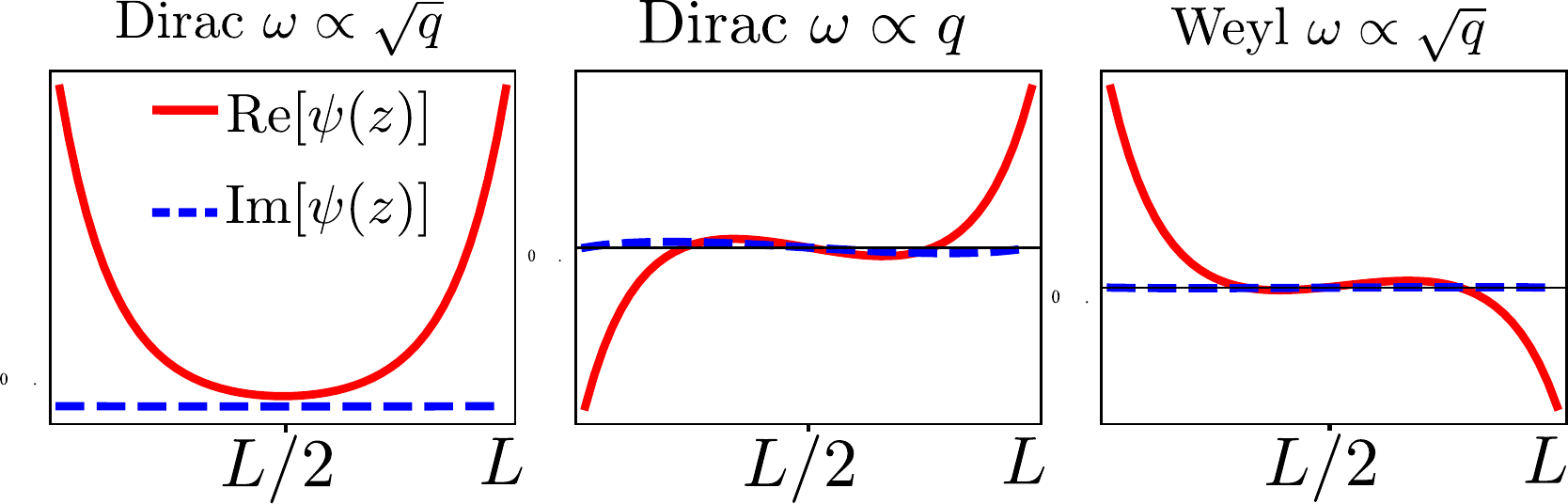}
	\caption{At the condition of the plasmonic mode, one of the eigenvalues of the matrix $I - \Delta z B(\mb{q},\omega)$ vanishes. We plot the corresponding (normalized) eigenvector $\psi(z)$, showing the $\omega \propto \sqrt{q}$ mode in the Weyl (right most) is indeed an anti-symmetric mode, which is contrary to the Dirac semimetal. Parameters are the same as in Fig.~\ref{fig:pldisp}.}
\label{fig:vec}
\end{figure}

Our numerical model allows one to construct the charge fluctuations associated with the collective modes~\cite{wvector} using the eigenvectors of the density response matrix $\chi({\bf q},z,z',\omega)$.  Results from such calculations are illustrated in Fig.~\ref{fig:vec}, and confirm the surprising difference between the WSM and the DSM systems: charge fluctuations which are antisymmetric across surfaces appear in a $\sqrt{q}$ mode for the WSM, whereas in the DSM -- as in conventional semiconductor bilayers -- this behavior is found in an acoustic mode.  Note that for similar parameter values, the antisymmetric mode of the WSM is considerably higher in frequency than the acoustic antisymmetric mode for the DSM, making the former more robust: the proximity of the acoustic mode to the particle-hole continuum edge makes it more susceptible to the broadening effects of disorder, which both relax momentum conservation and smear out the sharp edge of the continuum.


Another interesting aspect of the WSM plasmon modes is the evolution of their support on the surfaces as $q$ increases.  The heuristic model discussed above suggests their antisymmetry is eventually lost.  This is indeed the case, but rather than crossover into more standard symmetric behavior, we find that the modes become increasingly localized on one surface or the other, depending on the sign of $\mb{q}$ in the direction that the FA's disperse.  Thus the plasmons become similar to what would expect for excitations of a single FA.  The crossover between this latter behavior and the antisymmetric fluctuations occur around $q \sim 1/L$, where one expects interactions between surfaces to become important (see Appendix).

\section{Discussion}
In this study, we demonstrated that long-wavelength plasmons in a thin-film Weyl semimetal display the long-range nature of the Coulomb interaction by dispersing as $\sqrt{q}$, even as the associated charge oscillations are antisymmetric across surfaces.  This behavior contrasts with that of Dirac semimetals and conventional conducting bilayers, where such modes are symmetric.  This phenomenon is a direct result of the opposing chiralities of Fermi arc states on different surfaces.  The possibility of observing these modes is enhanced by the diverging slope as $q \rightarrow 0$, which keeps them well separated from the particle-hole continuum and the degrading effects this can have due to disorder effects.  Moreover, the dipole nature of the charge fluctuations suppresses fringing fields outside the thin film, which in practice can broaden these sharp modes, and limit their potential utility in plasmonic devices~\cite{landau}.  Interestingly, a dipole plasmon mode has very recently been observed \cite{Tanaka_2020}, albeit in a very different system, with very different underlying physics leading to the dipole nature of the mode.  Nevertheless, the  line-narrowing in the plasmon response due to suppression of fringing fields is indeed observed.

In currently available WSM's, surfaces typically support several FA's.  Interesting realizations of these are Co-based ferromagnets~\cite{CoSnS1,CoSnS2} which support three FA's on each surface, related by $2\pi/3$ rotations.  We expect thin films of such systems to support the antisymmetric plasmon modes we have studied here, although the modes are likely to be much less anisotropic with respect to wavevector.  Our studies suggest that thin films of this and other WSM materials are potential platforms for exotic low-dimensional plasmons, with behaviors that naturally emerge from their topological nature, making them unusually robust, and potentially useful in plasmonic systems.

\section{Acknowledgments}
A.K  acknowledges support from the SERB (Govt. of India) via saction no. ECR/2018/001443, DAE (Govt. of India ) via sanction no. 58/20/15/2019-BRNS, as well as MHRD (Govt. of India) via sanction no. SPARC/2018-2019/P538/SL. D.G  acknowledges the CSIR (Govt. of India) for financial support. H.A.F acknowledges support from the National Science Foundation via grant nos. ECCS-1936406 and DMR-1914451, as well as the support of the Research Corporation for Science Advancement through a Cottrell SEED Award, and the US-Israel Binational Science Foundation through award No. 2016130. We also  acknowledge the use of HPC facility at IIT Kanpur.

\renewcommand{\thefigure}{A\arabic{figure}}
\setcounter{figure}{0}
\renewcommand{\theequation}{A\arabic{equation}}
\setcounter{equation}{0}
\section*{Appendix}
\subsection*{A. Two layer systems}
\begin{figure}[t]
\centering
	\includegraphics[width=.45\textwidth]{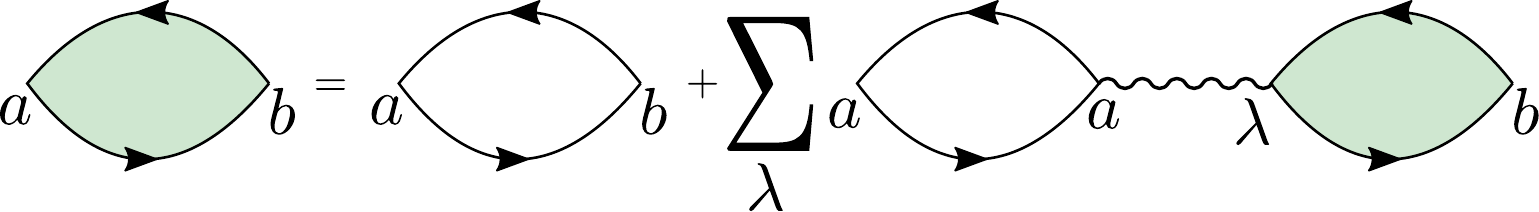}
	\caption{The self-consistent equations for the interacting response functions $\tilde{\chi}_{ab}$ (filled), where $a,b,\lambda=\pm1$ are layer indices, written at RPA approximation with single curly lines being the interaction $V_{a\lambda}$ and the unfilled loop being the non-interacting response function $\chi_{ab}\delta_{ab}$. The self-consistent equation reads $\tilde{\chi}_{ab} = \chi_{a}\delta_{ab} + \sum_{\lambda}\chi_{a}V_{a\lambda}\tilde{\chi}_{\lambda b}$. }\label{fig:FD}
\end{figure}

In this appendix we briefly review the formalism for collective modes in a bilayer system separated by a dielectric. Consider two two-dimensional (2D) systems with non-interacting polarization functions $\chi_{i}(\mb{q},\omega)$, and bare intra- and inter-layer Coulomb interactions given by 
$V_{ij}(\mb{q},\omega)$, with $i,j = 1,2$. Explicitly, $V_{11} = V_{22} = \alpha_c/q$ and $V_{12}$ = $V_{21} = \alpha_c e^{-qL}/q$, where $L$ is the separation between the layers and $\alpha_c = 2\pi e^2/\epsilon$. At the RPA level, if the interacting response functions are written as $\tilde{\chi}_{ij}(\mb{q},\omega)$, then (see Fig.~\ref{fig:FD}):
\begin{align}
\left(\begin{array}{cc}
1 - V_{11}\chi_{1} & -V_{12}\chi_1 \\
-V_{21}\chi_{2} & 1-V_{22}\chi_2
\end{array}\right)  \left(\begin{array}{c}
 \tilde{\chi}_{11} \\
  \tilde{\chi}_{21}
\end{array}\right)  = \left(\begin{array}{c}
 \chi_{1} \\
 0
\end{array}\right),
\end{align}
where for brevity, we have dropped the $\mb{q}, \omega$ indices. The matrix on the left is the dielectric matrix $\epsilon(\mb{q},\omega)$. An equivalent relation can be written for $\tilde{\chi}_{22}$ and $\tilde{\chi}_{12}$. Together, the interacting response matrix is written as
\begin{align}
\left(\begin{array}{cc}
\tilde{\chi}_{11} & \tilde{\chi}_{12} \\
\tilde{\chi}_{21} & \tilde{\chi}_{22}
\end{array}\right) = \epsilon^{-1}\left(\begin{array}{cc}
\chi_{1} & 0 \\
0 & \chi_{2}
\end{array}\right).
\end{align}

The conditions for collective modes are found from Det$~\epsilon(\mb{q},\omega)=0$, which gives  Eq.~(2) of the main text,
 \begin{align}\label{eq:slabcondApp}
& 1 - V(\mb{q})\big(\chi_1(\mb{q},\omega)
 + \chi_2(\mb{q},\omega) \big)\nonumber\\ &+V(\mb{q})^2(1-e^{-2qL}) \chi_1(\mb{q},\omega) \chi_2(\mb{q},\omega)=0.
\end{align}
In the limit when $L\ll q$, this equation reduces to
 \begin{align}
& \left(1 - V(\mb{q})\chi_1(\mb{q},\omega)\right)\left(1 - V(\mb{q})\chi_2(\mb{q},\omega)\right)=0,
\end{align}
which is the condition for decoupled collective modes for individual surfaces.

To analyze the nature of the charge oscillation for a plasmonic mode, it is useful to expand the $\chi_i$'s as a function of $q/\omega$ in the limit when $\omega > v_F q$, 
\begin{align}\label{eq:chiexp}
\chi_i = \mu_i\frac{ q}{\omega} + \nu_i\frac{ q^2}{\omega^2} + \cdots.
\end{align}
Furthermore, in the limit of small $q$, when $qL \ll 1$, one can write Eq.~(\ref{eq:slabcondApp}) as
 \begin{align}
& 1 - V\big(\chi_1 + \chi_2\big)\nonumber+2 qL ~V^2 \chi_1 \chi_2\approx0\nonumber\\
\Rightarrow & 1 - \left(\frac{a_1}{\omega} + \frac{a_2q}{\omega^2} + \cdots\right) + \left(\frac{b_1q}{\omega^2} + \cdots\right) \approx 0,
 \end{align}
Here $a_1 = \alpha_c(\mu_1+ \mu_2), a_2 = \alpha_c(\nu_1+\nu_2)$, and $b_1 = 2L\alpha_c^2\mu_1\mu_2$. Higher order terms in $q/\omega$ are unneeded in the analysis that follows. \\

\noindent
First, if $a_1$ is non-zero, then, the lowest order term in the above equation is of $q^0$, and the plasmon in the limit $\omega\gg v_Fq$ has a gap of order $\Delta =a_1$ in the limit $q\rightarrow 0.$\\

\noindent
If $a_1=0$ but $a_2\neq 0$, which as we explain below is the case of interest here, then to lowest order the equation becomes
\begin{align}
1 - \frac{a_2 q}{\omega^2} =0 \Rightarrow \omega =\sqrt{a_2-b_1}\sqrt{q}.
\end{align}
This is the $\sqrt{q}$ (optical) plasmon mode.  When $\omega$ and $q$ satisfy this dispersion relation the determinant of the dielectric function vanishes.


\subsubsection*{Polarizability of the Fermi Arcs}
 The energy dispersion of the Fermi arc (FA) states on the top and bottom surfaces of a thin-film Dirac semimetal is modeled by
 \begin{align}
 E_s^{(\pm)}(\mb{k})=(\pm) s \hbar v_F  k_x;~s=\pm 1;~-k_0 < k_y < k_0,
 \label{Epm}
 \end{align}
where the (pseudo-)spin index $s=1$ on one surface and $s=-1$ on the other for Weyl FA's, whereas they both are present on each of the surfaces for the Dirac FA. At charge neutrality, the Fermi arcs are straight-lines joining the Weyl nodes from $k_y=-k_0$ to $k_y=k_0$. In the limit of small $q$, for a given spin-sector $s=1$, for the top surface (+ sign in Eq. \ref{Epm}),
 \begin{align}
 \chi_{1,s=1}(\mb{q},\omega)&=\int \frac{d^2\mb{k}}{4\pi^2}
  \frac{n_F(E_\mb{k})-n_F(E_{\mb{k}+\mb{q}})}{\hbar \omega -\hbar v_F q_x}\nonumber \\
 &=-\int \frac{d^2\mb{k}}{4\pi^2}~ \frac{\partial n_F(E_\mb{k})}{\partial E_\mb{k}} \mb{q}.\frac{\partial E_\mb{k}}{\partial \mb{k}} ~\frac{1}{\hbar \omega -\hbar v_F q_x}\nonumber\\
 &=\int \frac{d^2\mb{k}}{4\pi^2} ~\frac{\hbar v_F q_x \delta(E_\mb{k}-E_F)}{\hbar \omega -\hbar v_F q_x}\nonumber\\
 &=\int \frac{dk_y dk_x}{4\pi^2}~\frac{q_x\delta(k_x-k_F)}{\hbar \omega -\hbar v_F q_x}\nonumber\\
 &=\frac{k_0}{2\pi^2} \frac{q_x}{\hbar \omega -\hbar v_F q_x}\label{eq:chi0s1}.
 \end{align}
  We write $\omega=v_F k_F \bar{\omega}$, $q_x=k_F \bar{q}_x$ and $k_0=k_F \bar{k}_0$, so that $\bar{\omega}$, $\bar{q}_x$ and $\bar{k}_0$ become dimensionless variables. Then Eq.~(\ref{eq:chi0s1}) may be written in the form
  \begin{align}
  \chi_{1,s=1}(\mb{q},\omega)&\equiv \chi_1^{\rm WSM}(\mb{q},\omega) = \frac{k_F}{2\pi^2\hbar v_F } \frac{\bar{k}_0 \bar{q}_x}{\bar{\omega} -\bar{q}_x}\nonumber\\
  &=\frac{\beta\bar{q}_x}{\bar{\omega} -\bar{q}_x}\label{eq:chi0s2},
  \end{align}
  where $\beta=\frac{k_F\bar{k}_0}{2\pi^2\hbar v_F}$.
  Similarly,
  \begin{align}
 \chi_{1,s=-1}(\mb{q},\omega) \equiv \chi_2^{\rm WSM}(\mb{q},\omega) =-\frac{\beta\bar{q}_x}{\bar{\omega} +\bar{q}_x}\label{eq:chi0ms}.
  \end{align}
  The non-interacting polarizabilities for the top and bottom surfaces of a Dirac semimetal are same as one another and are given by
  \begin{align}
  \chi^{\rm DSM}_1(\mb{q},\omega)=\chi^{\rm DSM}_2(\mb{q},\omega)&=\beta\big( \frac{\bar{q}_x}{\bar{\omega} -\bar{q}_x} - \frac{\bar{q}_x}{\bar{\omega} +\bar{q}_x}      \big)\nonumber\\
  &=\frac{2\beta\bar{q}_x^2}{\bar{\omega}^2-\bar{q}_x^2}.\label{eq:chi0T}
  \end{align}
These are the expressions used in the main text. Expanding for small $\bar{q}/\bar{\omega}$, similar to Eq.~(\ref{eq:chiexp}), one finds $\mu_1 = -\mu_2 = \beta \cos\theta$, $\nu_1 = \nu_2 = \beta \cos^2\theta$ for the Weyl FA. For the Dirac FA, $\mu_i=0$ and $\nu_1 = \nu_2 = 2\beta \cos^2\theta$. (In these expressions $\theta = \cos^{-1}\frac{q_x}{q}$.)

\subsubsection*{Single-surface plasmon modes}
For a single surface with Dirac or Weyl FA, the dispersions of the plasmon modes can be found by solving 
\begin{align}
1- V(q)\chi(\mb{q},\omega)=0.
\end{align}
For the Dirac FA,the equation reduces to:
\begin{align}
\bar{\omega}^2-\bar{q}^2\cos^2\theta - 2\bar{\alpha}_c\beta\bar{q}\cos^2\theta=0,\nonumber
\end{align}
where $\bar{\alpha}_c = \alpha_c/k_F$. When $\bar{q}\ll \bar{\omega}$, this results in a single plasmon mode with dispersion
\begin{align}
\omega = v_F\sqrt{2\alpha_c\beta}\cos\theta \sqrt{q}.
\end{align}
For the Weyl FA, the same equation reduces to
\begin{align}
\bar{\omega}-s\bar{q}\cos\theta - \bar{\alpha}_c\beta\cos\theta=0,\nonumber
\end{align}
resulting in a gapped, chiral plasmon mode
\begin{align}
\omega = v_F\alpha\beta\cos\theta + sv_F\cos\theta q.\label{eq:singleWFA}
\end{align}
The chirality of the plasmon mode is the result of the chirality of the FA states.

\subsubsection*{Two-surface plasmon modes}
For the two-surfaces of the slab-geometry, we substitute the non-interacting response functions of the two surfaces in Eq.~(\ref{eq:slabcondApp}) for the collective modes. For the Dirac system, as $\chi_1^{{\rm DSM}} = \chi_2^{{\rm DSM}} = \chi^{{\rm DSM}}$, the equation reduces to
\begin{align}
1 - V(q)\chi^{{\rm DSM}}(\mb{q},\omega) = \pm e^{-qL}\chi^{{\rm DSM}}(\mb{q},\omega).
\end{align}
For the (+) on the right-hand-side, for $qL\ll 1$, this results in  the dispersion
\begin{align}
\bar{\omega}^2 - \cos^2\theta\bar{q}^2 - 4\bar{\alpha}_c\beta\cos^2\theta\bar{q}=0,
\end{align}
For the (-) sign, for $qL\ll 1$, we obtain the dispersion
\begin{align}
\bar{\omega}^2 -  \cos^2\theta\bar{q}^2 - 2\bar{\alpha}_c\beta \bar{L}\cos^2\theta\bar{q}^2=0,
\end{align}
where $\bar{L} = Lk_F$.  Keeping smallest orders in $\bar{q}$, thus we get two plasmon modes with dispersions
 \begin{align}
 \omega^{(1)}_{\rm D} = v_F\sqrt{1+2\alpha_c\beta L}\cos\theta ~q, ~~ \omega^{(2)}_{\rm D} = v_F \sqrt{4\alpha_c\beta}\cos\theta ~\sqrt{q}.\label{eq:pld}
\end{align}
By contrast, for the WSM, in the limit of $qL\ll1$, the equation for collective mode reduces to
\begin{align}
\bar{\omega} - \bar{q}^2\cos^2\theta \approx 2\bar{\alpha}_c\beta(1+\bar{\alpha}_c\beta\bar{L})\cos^2\theta \bar{q}.
\end{align}
In the lowest order in $\bar{q}$, we obtain the plasmon dispersion (in terms of dimension full variables) 
\begin{align}
\omega_{\rm W} =v_F \sqrt{2\alpha_c\beta}\sqrt{1 + \alpha_c\beta L}\cos\theta \sqrt{q}.\label{eq:plw}
\end{align}
Notice that in either case the plasmon dispersions become steeper with increasing $L,k_0$ and become softer with increasing $\theta$ from $0^{{\rm o}}$.  We have verified these behaviors numerically.

\subsubsection*{Charge oscillation for the $\sqrt{q}$ mode}
The net charge fluctuations on the two surfaces $\delta \rho_i(\mb{q},\omega)$ ($i=1,2$) can be written in terms of the response functions in the presence of external potentials $\phi_{{\rm ext},i}$ on the $i$th surface as
\begin{align}
&\left(\begin{array}{c}
\delta\rho_1 \\
\delta\rho_2
\end{array}\right) = \left(\begin{array}{cc}
\tilde{\chi}_{11} & \tilde{\chi}_{12} \\
\tilde{\chi}_{21} & \tilde{\chi}_{22}
\end{array}\right) \left(\begin{array}{c}
\phi_{{\rm ext}, 1}\\
\phi_{{\rm ext}, 2}
\end{array}\right) = \epsilon^{-1}\left(\begin{array}{c}
\chi_1\phi_{{\rm ext}, 1}\\
\chi_2\phi_{{\rm ext}, 2}
\end{array}\right)\nonumber\\
\Rightarrow~ & \epsilon \left(\begin{array}{c}
\delta\rho_1 \\
\delta\rho_2
\end{array}\right) = \left(\begin{array}{c}
\chi_1\phi_{{\rm ext}, 1}\\
\chi_2\phi_{{\rm ext}, 2}
\end{array}\right).
\end{align}
For self-sustained charge oscillations, $(\delta\rho_1,\delta\rho_2)$ is the eigenvector of the $\epsilon$ matrix with zero eigenvalue. This implies
\begin{align}
\frac{\delta\rho_1}{\delta\rho_2} = \frac{\chi_1V_{12}}{1-V_{11}\chi_1}.
\end{align}

For the WSM, $\mu_i\neq0$, so that for small $q$ and for the $\omega\propto\sqrt{q}$ mode, $\chi_1\propto \sqrt{q}$ whereas $V_{11},V_{12}\propto 1/q$.  This implies
\begin{align}
\left.\frac{\delta\rho_1}{\delta\rho_2}\right|_{{\rm WSM}} \approx -\frac{V_{12}}{V_{11}} \approx -1,
\end{align}
resulting in antisymmetric oscillation. We note that this anti-symmetric nature also holds for the eigenvector of the $\epsilon$ matrix when $\omega$ and ${\bf q}$ satisfy the plasmon mode dispersion relation.

For the case of a Dirac semimetal (as well as for a normal metal), for small $q$ and for the $\omega \propto \sqrt{q}$ mode, $\chi_i\propto q$. In this case, the resulting density amplitudes follow
\begin{align}
\left.\frac{\delta\rho_1}{\delta\rho_2}\right|_{{\rm Dirac}} \approx \frac{\nu_1}{\nu_2}.
\end{align}
As $\nu_1=\nu_2$, this implies a symmetric charge oscillation. For the $\omega\propto q$ mode, a similar argument yields $\delta\rho_1/\delta\rho_2 \approx -V_{12}/V_{11}$, i.e, an antisymmetric charge-oscillation mode.

\subsection*{B. Eigenstates in slab geometry}
The low-energy Hamiltonian we consider in the main text, which contains two Weyl nodes labeled by $\eta=\pm1$, is
\begin{align}\label{eq:newH}
H_{\eta} = (\sigma_y q_x - \sigma_x q_z) + \sigma_z M_{\eta}(k_y),
\end{align}
with $M_{\eta}(k_y) = \eta (m - \cos k_y)$. The two Weyl nodes are at
$\mb{k} = (0,\pm k_0,0)$ with $k_0 = \cos^{-1}(m/\lambda)$.
For the $\eta=+1$ block, $M_+ < 0 $ between $k_y\in (-k_0,k_0)$.
For a surface perpendicular to the $z$ direction, along the $k_y$ axis
these two points are connected by a Fermi arc on the surface Brillouin zone.  For the $\eta=-1$ block, $M_- > 0 $ between $k_y \in (-k_0,k_0)$, and again there is a Fermi arc connecting these points on the $k_y$ axis for the same surface. The WSM/DSM slab is confined between $z=0$ and $z=L$.

Following ref.~20 of the main text, we infinite mass boundary conditions by taking the Hamiltonian of the vacuum to be same as Eq.~(\ref{eq:newH}), except for the mass term, whose form is taken to be $M^{\rm vac}_{\eta} = \eta m_0$, with $m_0 \rightarrow \infty$. This construction is required to ensure that for $k_y$ between the Weyl nodes the effective mass term ($M_{\eta}(k_y)$) for the Weyl semimetal and the vacuum ($M_{\eta}^{\rm vac}$) are oppositely signed. By matching the wave-function at the boundary one arrives at the transcendental equation
\begin{align}
&\frac{{\rm tanh}\left(L\sqrt{M_{\eta}^2 - \xi}\right)}{L\sqrt{M_{\eta}^2 -\xi}} = -\frac{1}{L\eta M_{\eta}}.
\label{eq:chi}
\end{align}
Solutions of this equation, $\xi_n$, which we label by the band-index $n$, yields the band energies $E_n = {\rm sign}(n)\sqrt{\xi_n+q_x^2}$.

For a given solution of energy $E$, for the block $\eta$, defining $K = M_{\eta}(k_y) -E, f=q_x-iq_z, g=q_x+iq_z$, one finds the corresponding wave-functions
\begin{align}\label{eq:wf1}
\phi_{\eta,\mb{k}}(z) =& \frac{1}{\sqrt{N}} \left\{ (K+\eta g)\left(\begin{array}{c}
i f \\
K
\end{array} \right)e^{iq_z z}\right.\nonumber\\
& \left. ~~~~~~~~~~~~~+(K+\eta f)\left(\begin{array}{c}
-i g \\
-K
\end{array} \right)e^{-iq_z z} \right\}.
\end{align}
For real $q_z = \sqrt{\chi - M_{\eta}(k_y)^2}$ (when $\chi>m^2$, $f=g^*$) the normalization factor has the form
\begin{align}
N =& 2|K+\eta f|^2(K^2+|f|^2)L\nonumber \\
&+{\rm Im}\left[(K+\eta f)^2(K^2+g^2)\left(\frac{e^{-2iLq_z}-1}{q_z}\right) \right].
\end{align}
For purely imaginary $q_z=i\kappa$ (when $\chi<m^2$), $f=q_x+\kappa$, $g=q_x-\kappa$, the normalization is
\begin{align}
N =&-2(K+\eta f)(K+\eta g)(K^2+g f) L\nonumber \\
&+[(K+\eta g)^2(f^2+K^2)e^{-\kappa L}\nonumber\\
&+(K+\eta f)^2(g^2+K^2) e^{\kappa L}]\frac{\sinh(\kappa L)}{\kappa}.
\end{align}
These are the full solutions of the low-energy states for the semimetal slab.

\begin{widetext}
\subsection*{C. Details of the density-density response function}
In terms of the wave-functions, we write the charge density operator, $ \rho (\mb{r},z)= \Psi (\mb{r},z)^{\dagger} \Psi (\mb{r},z)$ where $\mb{r}=(x,y)$ and $\Psi(\mb{r},z)$ is the field operator	
	\begin{align}\label{eq:fod}
	\Psi(\mb{r},z)=\frac{1}{\sqrt{L_x L_y}}\sum_{\mb{k}} \exp(i \mb{r}.\mb{k}) \sum_{\eta,m,} \phi_{\eta,m,\mb{k}} (z) c_{\eta,m,\mb{k}},\nonumber
	\end{align}
where the summation over the band indices $m$ include positive as well as negative bands. $c_{\eta,m,\mb{k}}$ is the electronic annihilation operator and the $\eta$ summation is absent in the case of WSM. $\phi_{\eta,m,\mb{k}}$ are eigenstates of the non-interacting Hamiltonian, i.e, $H_{0}(\mb{k})\phi_{\eta,m,\mb{k}} = \epsilon_{\eta,m,\mb{k}}\phi_{\eta,m,\mb{k}}$. The interaction Hamiltonian is then
\begin{align}
H_{\rm int}	&=\frac{1}{2}\int d\mb{R}_1 d\mb{R}_2 V_c(|\mb{R}_1-\mb{R}_2|) :\rho(\mb{R}_1) \rho(\mb{R}_2):,\nonumber
\end{align}
where $\mb{R} = (\mb{r},z)$. Using the Fourier transformed form $V(|\mb{R}_1-\mb{R}_2|)=(1/4\pi^2)\int d^2 \mb{q}_1 V_{\mb{q}_1}(|z_1-z_2|)$, where $V_{\mb{q}}(|z_1-z_2|) = 2\pi\alpha \exp(-q|z_1-z_2|)/q $, $H_{\text{int}}$ can be written as
\begin{align}
H_{\text{int}}=\frac{1}{8\pi^2} \int_{0}^{L} \int_{0}^{L} dz_1 dz_2 \int d^2 \mb{q}_1 V_{\mb{q}_1}(|z_1-z_2|)\sum_{ \eta_1,\eta_2, \{l_i\},\mb{k}_1,\mb{k}_2} \phi^\dagger_{\eta_1,l_1,\mb{k}_1}(z_1)~\phi_{\eta_1,l_4,\mb{k}_1-\mb{q}_1}(z_1)~\phi^\dagger_{\eta_2,l_2,\mb{k}_2}(z_2)~\phi_{\eta_2,l_3,\mb{k}_2+\mb{q}_1}(z_2)\nonumber \\  \times ~c^\dagger_{\eta_1,l_1,\mb{k}_1}
~c^\dagger_{\eta_2,l_2,\mb{k}_2}~c_{\eta_2,l_3,\mb{k}_2+\mb{q}_1}~c_{\eta_1,l_4,\mb{k}_1-\mb{q}_1}.\nonumber
\end{align}

 The time-dependent density-density response function is defined as
\begin{align}
\chi({\bf q},z,z', t)&=
-\textit i  \theta (t)\int d\mb{r}e^{i{\bf q} \cdot {\bf r}}
\langle  \big[\rho(\mb{r},z,t), \rho(0,z',0)] \big\rangle\nonumber\\
&\equiv \frac{1}{L_x L_y}\sum_{\eta,\eta^{\prime},\mb{k},\mb{k}^{\prime},m,m^{\prime},s,s^{\prime}} \chi_{\eta,\eta^{\prime},m,m^\prime,s,s^\prime}(\mb{k},\mb{k}^\prime,\mb{q},z,z^\prime,t),
\end{align}
where
\begin{align}\nonumber
\chi_{\eta,\eta^{\prime},m,m^\prime,s,s^\prime}(\mb{k},\mb{k}^\prime,\mb{q},z,z^\prime,t)=&-i\theta(t)~\phi^{\dagger}_{\eta,m,\mb{k}} (z)~ \phi_{\eta,m^{\prime},\mb{k}-\mb{q}} (z) ~\phi^{\dagger}_{\eta^{\prime},s,\mb{k}^{\prime}} (z^{\prime}) ~\phi_{\eta^{\prime},s^{\prime},\mb{k}^{\prime}-\mb{q}} (z^{\prime})  \\
&\times	\big\langle \big[ \exp(i  H t)~c^{\dagger}_{\eta,m,\mb{k}^{\prime}}~ c_{\eta,m^{\prime},\mb{k}+\mb{q}}~ \exp(-i  H t),  ~c^{\dagger}_{\eta^{\prime},s,\mb{k}^{\prime}} ~c_{\eta^{\prime},s^{\prime},\mb{k}^{\prime}-\mb{q} }~\big]\big \rangle. \label{eq:chid}
\end{align}
 We take time derivative of Eq.~(\ref{eq:chid}) to arrive at
\begin{align}\nonumber
\partial_t ~\chi_{\eta,\eta^{\prime},m,m^\prime,s,s^\prime}(\mb{k},\mb{k}^\prime,\mb{q},z,z^\prime,t)= -i \delta(t)~\phi^{\dagger}_{\eta,m,\mb{k}} (z)~ \phi_{\eta,m^{\prime},\mb{k}-\mb{q}} (z) ~\phi^{\dagger}_{\eta^{\prime},s,\mb{k}^{\prime}} (z^{\prime}) ~\phi_{\eta^{\prime},s^{\prime},\mb{k}^{\prime}-\mb{q}} (z^{\prime})\nonumber\\
\times \big\langle \big[~c^{\dagger}_{\eta,m,\mb{k}^{\prime}}~ c_{\eta,m^{\prime},\mb{k}+\mb{q}}~,  ~c^{\dagger}_{\eta^{\prime},s,\mb{k}^{\prime}} ~c_{\eta^{\prime},s^{\prime},\mb{k}^{\prime}-\mb{q} }~\big]\big \rangle \nonumber\\
-i\theta(t)~\phi^{\dagger}_{\eta,m,\mb{k}} (z)~ \phi_{\eta,m^{\prime},\mb{k}-\mb{q}} (z) ~\phi^{\dagger}_{\eta^{\prime},s,\mb{k}^{\prime}} (z^{\prime}) ~\phi_{\eta^{\prime},s^{\prime},\mb{k}^{\prime}-\mb{q}} (z^{\prime}) \nonumber \\
\times	\big\langle i \big[ \exp(i  H t)~[H,c^{\dagger}_{\eta,m,\mb{k}^{\prime}}~ c_{\eta,m^{\prime},\mb{k}+\mb{q}}]~ \exp(-i  H t),  ~c^{\dagger}_{\eta^{\prime},s,\mb{k}^{\prime}} ~c_{\eta^{\prime},s^{\prime},\mb{k}^{\prime}-\mb{q} }~\big]\big \rangle. \label{eq:chid1}
\end{align}
The commutators of the single particle terms are easily evaluated, yielding
\begin{align}
&\big [c^{\dagger}_{\eta,m,\mb{k}}c_{\eta,m^{\prime},\mb{k}+\mb{q}} ,~ c^{\dagger}_{\eta^{\prime},s,\mb{k}^{\prime}}c_{\eta^{\prime},s^{\prime},\mb{k}^{\prime}-\mb{q}}  \big]=\big(n_F(\epsilon_{\eta,m,\mb{k}}) - n_F(\epsilon_{\eta,m^{\prime},\mb{k}+\mb{q}})~\big)\delta_{\eta,\eta^{\prime}} \delta_{m,s^{\prime}}\delta_{m^{\prime},s} \delta_{\mb{k}+\mb{q},\mb{k}^{\prime}},\label{eq:t1}\\
&\big[H_0, c^{\dagger}_{\eta,m,\mb{k}}c_{\eta,m^{\prime},\mb{k}+\mb{q}}\big]=\big( \epsilon_{\eta,m,\mb{k}} -\epsilon_{\eta,m^{\prime},\mb{k}+\mb{q}} \big)c^{\dagger}_{\eta,m,\mb{k}}c_{\eta,m^{\prime},\mb{k}+\mb{q}}\label{eq:t2}.
\end{align}
For the interaction term we use the Hartree approximation, so that one makes the replacement
 \begin{align}
\big[H_{\text{int}},c^{\dagger}_{\eta,m,\mb{k}} c_{\eta,m^{\prime},\mb{k}+\mb{q}}\big]~
& \rightarrow ~\big[ n_F(\epsilon_{\eta,m^{\prime},\mb{k}+\mb{q}})- n_F(\epsilon_{\eta',m,\mb{k}})\big]\int_{0}^{L} \int_{0}^{L} dz_1 dz_2  V_{\mb{q}_1}(|z_1-z_2|)\nonumber\\ &\times \phi^\dagger_{\eta,m^{\prime},\mb{k}+\mb{q}}(z_1)~\phi_{\eta,m,\mb{k}}(z_1)\sum_{\eta_2,l_2,l_3,\mb{k}_2} \phi^\dagger_{\eta_2,l_2,\mb{k}_2}(z_2)~\phi_{\eta_2,l_3,\mb{k}_2+\mb{q}}(z_2)~c^\dagger_{\eta_2,l_2,\mb{k}_2}c_{\eta_2,l_3,\mb{k}_2+\mb{q}}\label{eq:HAf2}.
\end{align}
Using Eq.~(\ref{eq:t1})-(\ref{eq:HAf2}) in Eq.~(\ref{eq:chid1}) leads to the self-consistent equation
\begin{align}\nonumber
\partial_t \chi_{\eta,\eta^{\prime},m,m^\prime,s,s^\prime}(\mb{k},\mb{k}^\prime,\mb{q},z,z^\prime,t)= &-i \delta(t)F_{\eta,m,m^\prime}(\mb{k},\mb{q},z,z^\prime)~\big[~n_F(\epsilon_{\eta,m,\mb{k}}) - n_F(\epsilon_{\eta,m^{\prime},\mb{k}+\mb{q}})~\big]~\delta_{\eta,\eta^{\prime}} ~ \delta_{m,s^{\prime}}~\delta_{m^{\prime},s}~ \delta_{\mb{k}+\mb{q},\mb{k}^{\prime}}   \nonumber\\
&+i~(\epsilon_{\eta,m,\mb{k}}-\epsilon_{\eta,m^\prime,\mb{k}+\mb{q}})\chi_{\eta,\eta^{\prime},m,m^\prime,s,s^\prime}(\mb{k},\mb{k}^\prime,\mb{q},z,z^\prime,t)\nonumber\\
&+i\big[n_F(\epsilon_{\eta,m^{\prime},\mb{k}+\mb{q}})-n_F(\epsilon_{\eta',m,\mb{k}})\big]\int_{0}^{L} \int_{0}^{L} dz_1 dz_2 V_{\mb{q}_1}(|z_1-z_2|)F_{\eta,m,m^\prime}(\mb{k},\mb{q},z,z_1)\nonumber\\
&~~~~~~~~~~~~~~~~~~~~~~~~~~~~~~~~~~~~~~~~~~~~\times\sum_{\eta_2,l_2,l_3,\mb{k}_2}\chi_{\eta_2,\eta^{\prime},l_2,l_3,s,s^\prime}(\mb{k}_2,\mb{k}^\prime,\mb{q},z_2,z^\prime,t)\label{eq:chid2},
\end{align}
where
\begin{align}
F_{\eta,m,m^\prime}(\mb{k},\mb{q},z,z^\prime)&=\phi^{\dagger}_{\eta,m,\mb{k}} (z)~ \phi_{\eta,m^{\prime},\mb{k}+\mb{q}} (z) ~\phi^{\dagger}_{\eta,m^\prime,\mb{k}+\mb{q}} (z^{\prime}) ~\phi_{\eta,m,\mb{k}} (z^{\prime}).
\end{align}
Fourier transforming Eq.~(\ref{eq:chid2}) with respect to time, this equation may be recast as
\begin{align}\nonumber
-i\omega~ \chi_{\eta,\eta^{\prime},m,m^\prime,s,s^\prime}(\mb{k},\mb{k}^\prime,\mb{q},z,z^\prime,\omega)=&-iF_{\eta,m,m^\prime}(\mb{k},\mb{q},z,z^\prime)~\big[~n_F(\epsilon_{\eta,m,\mb{k}}) - n_F(\epsilon_{\eta,m^{\prime},\mb{k}+\mb{q}})~\big]~\delta_{\eta,\eta^{\prime}} ~ \delta_{m,s^{\prime}}~\delta_{m^{\prime},s}~ \delta_{\mb{k}+\mb{q},\mb{k}^{\prime}}   \nonumber\\
&+i~(\epsilon_{\eta,m,\mb{k}}-\epsilon_{\eta,m^\prime,\mb{k}+\mb{q}})\chi_{\eta,\eta^{\prime},m,m^\prime,s,s^\prime}(\mb{k},\mb{k}^\prime,\mb{q},z,z^\prime,t)\nonumber\\
&+i\big[n_F(\epsilon_{\eta,m^{\prime},\mb{k}+\mb{q}})-n_F(\epsilon_{\eta',m,\mb{k}})\big]\int_{0}^{L} \int_{0}^{L} dz_1 dz_2 V_{\mb{q}_1}(|z_1-z_2|)F_{\eta,m,m^\prime}(\mb{k},\mb{q},z,z_1)\nonumber\\
&~~~~~~~~~~~~~~~~~~~~~~~~~~~~~~~~~~~~~~~~~~~~\times\sum_{\eta_2,l_2,l_3,\mb{k}_2}\chi_{\eta_2,\eta^{\prime},l_2,l_3,s,s^\prime}(\mb{k}_2,\mb{k}^\prime,\mb{q},z_2,z^\prime,\omega)\label{eq:chid3}.
\end{align}
\begin{figure*}
\centering
	\includegraphics[width=.95\textwidth]{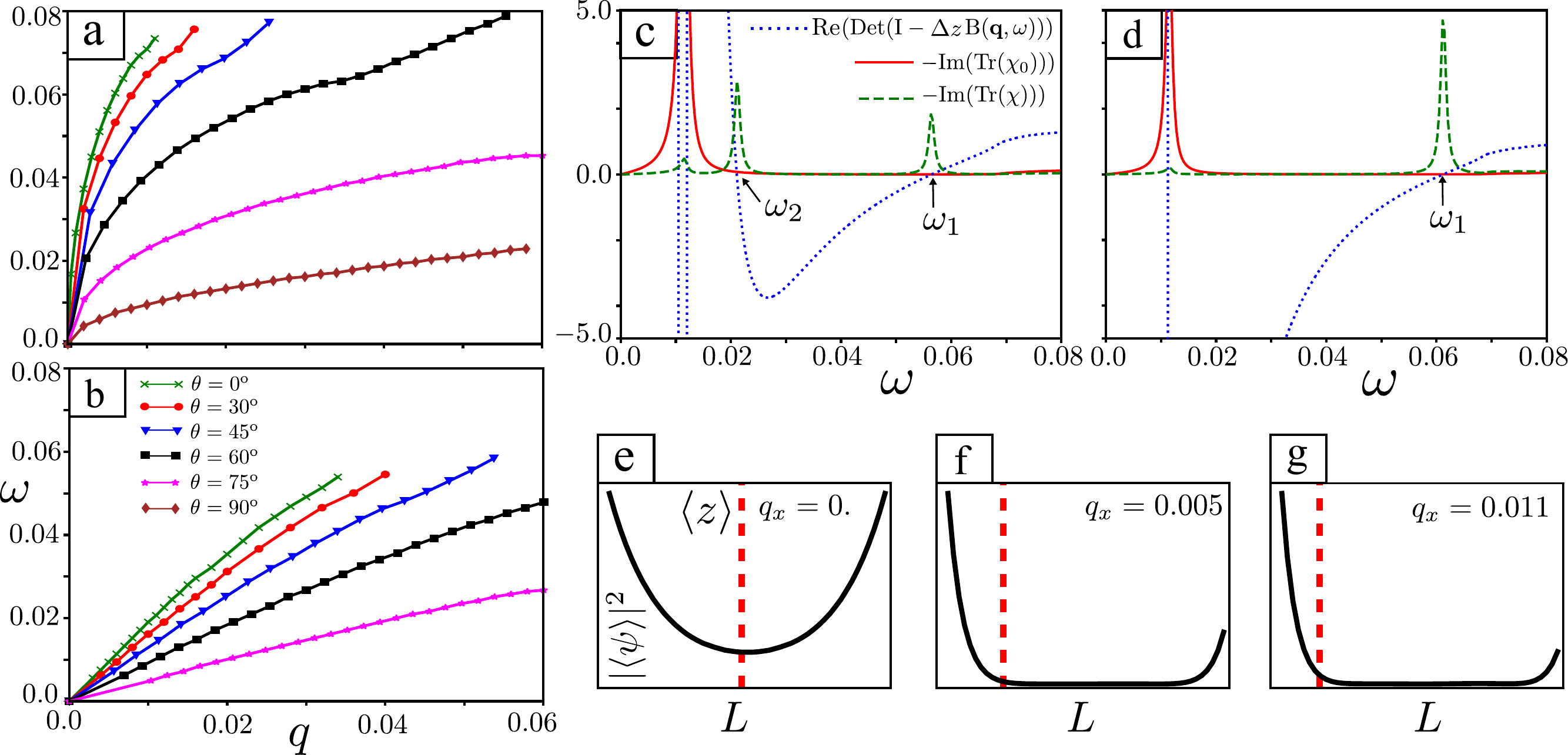}
	\caption{(a), (b): The variation of the two plasmon dispersions, for the $\omega\propto\sqrt{q}$ and $\omega\propto q$ modes, respectively, as a function of the angle $\theta=\cos^{-1}q_x/q$, for the case of the Dirac system (for Weyl system the variation for the single plasmon mode is qualitatively similar to (a)). At $\theta=90^{{\rm o}}$ we obtain a single mode for the Dirac system. (c), (d): Nature of the response function for  (c) Dirac and (d) Weyl system. Sharp collective modes appear for two values of $\omega$ for a given $q$ in case of the Dirac system and once for Weyl system. Broader modes with Im$\chi_0\neq0$ here represent the intra-band particle-hole continuum.  (e)-(g) We plot the eigen-vector of the plasmon distribution, $\psi(z)$ (see Sec IV) and indicate the mean value of $z$ in this eigen-vector $\langle z\rangle = \sum_{i=0}^{n}\Delta z z_i \psi(z_i)$, where $i$ denotes the discretization of the $z$ direction between $0$ and $L$ in $n$ parts with $\Delta z = L/n$. For (c)-(g), parameters chosen are the same as in Fig.~2 of the main text with $\theta=60^{{\rm o}}$, except in Fig. (e-g), $\theta=0^{{\rm o}}$.}\label{fig:wtheta}
\end{figure*}
Summing over the indices $\eta$, $\eta^{\prime}$ $m$, $m^\prime$, $s$, $s^\prime$, $\mb{k}$, $\mb{k}^\prime$, we obtain
\begin{align}\nonumber
\chi(\mb{q},z,z^{\prime},\omega)&=\frac{1}{L_x L_y} \sum_{\eta,\mb{k},m,m^{\prime}}  \frac{ n_F(\epsilon_{\eta,m,\mb{k}}) - n_F(\epsilon_{\eta,m^{\prime},\mb{k}+\mb{q}}) }{\omega + i \eta + \epsilon_{\eta,m,\mb{k}} - \epsilon_{\eta,m^{\prime},\mb{k}+\mb{q}} } \bigg[ F_{\eta,m,m^{\prime}}(\mb{k},\mb{q},z,z^{\prime})\nonumber \\
&+ \int_{0}^{L}\int_{0}^{L} dz_1 dz_2 V_{\mb{q}}(| z_1-z_2|) F_{\eta,m,m^\prime}(\mb{k},\mb{q},z,z_1) \chi(\mb{q},z_2,z^{\prime},\omega) \bigg]\nonumber\\
&=\chi_0(\mb{q},z,z^{\prime},\omega) + \int_{0}^{L} \int_{0}^{L} dz_1 dz_2 V_{\mb{q}}(| z_1-z_2|) \chi_0(\mb{q},z,z_1,\omega) \chi(\mb{q},z_2,z^{\prime},\omega),\label{eq:finald}
\end{align}
where the non-interacting response function has the form
\begin{align}\label{eq:chi0d}
\chi_0(\mb{q},z,z^{\prime},\omega)=\frac{1}{L_x L_y}\sum_{\eta,\mb{k},m,m^{\prime}} \frac{ n_F(\epsilon_{\eta,m,\mb{k}}) - n_F(\epsilon_{\eta,m^{\prime},\mb{k}+\mb{q}}) }{\omega + i \eta + \epsilon_{\eta,m,\mb{k}} - \epsilon_{\eta,m^{\prime},\mb{k}+\mb{q}} } F_{\eta,m,m^{\prime}}(\mb{k},\mb{q},z,z^{\prime}).
\end{align}
The integration over $z_1$ can be performed analytically in Eq.~(\ref{eq:finald}), allowing it to be rewritten as
\begin{align}\label{eq:final2}
\chi(\mb{q},z,z^{\prime},\omega)= \chi_0(\mb{q},z,z^{\prime},\omega) + \int_{0}^{L}  dz_2  B(\mb{q},z,z_2,\omega) \chi(\mb{q},z_2,z^{\prime},\omega) ,
\end{align}
with
\begin{align}
B(\mb{q},z,z_2,\omega) = \int_{0}^{L} dz_1  V_{\mb{q}}(| z_1-z_2|) \chi_0(\mb{q},z,z_1,\omega).
\end{align}
We convert the integration over $z_2$ in Eq.~(\ref{eq:final2}) into a summation over $N$ discrete $z_2$ points, allowing us to arrive at
\begin{align}\label{eq:final3}
\chi(\mb{q},z_i,z_j,\omega)= \chi_0(\mb{q},z_i,z_j,\omega) + \sum_{k=1}^{N}  \Delta z  B(\mb{q},z_i,z_k,\omega) \chi(\mb{q},z_k,z_j,\omega),
\end{align}
with $\Delta z= L/(N-1)$ and each $z_i$ are now discretized with $i=1,\cdots,N$.
\end{widetext}
Eq.~(\ref{eq:final3}) can alternatively be written in matrix form,
\begin{align}
\chi(\mb{q},\omega) &= \chi_0(\mb{q},\omega) +   \Delta z  B(\mb{q},\omega) \chi(\mb{q},\omega)\nonumber\\
\Rightarrow ~ \chi(\mb{q},\omega) &= ( I-\Delta z B(\mb{q},\omega) )^{-1}\chi_0(\mb{q},\omega).
\end{align}
The entire calculation is similar for Dirac and Weyl semimetals, except that there is no summation over $\eta$ for the case of the Weyl system.
The condition for plasmon modes then reads ${\rm det}[ I-\Delta z B(\mb{q},\omega) ]=0$.
\begin{figure*}
\centering
	\includegraphics[width=.9\textwidth]{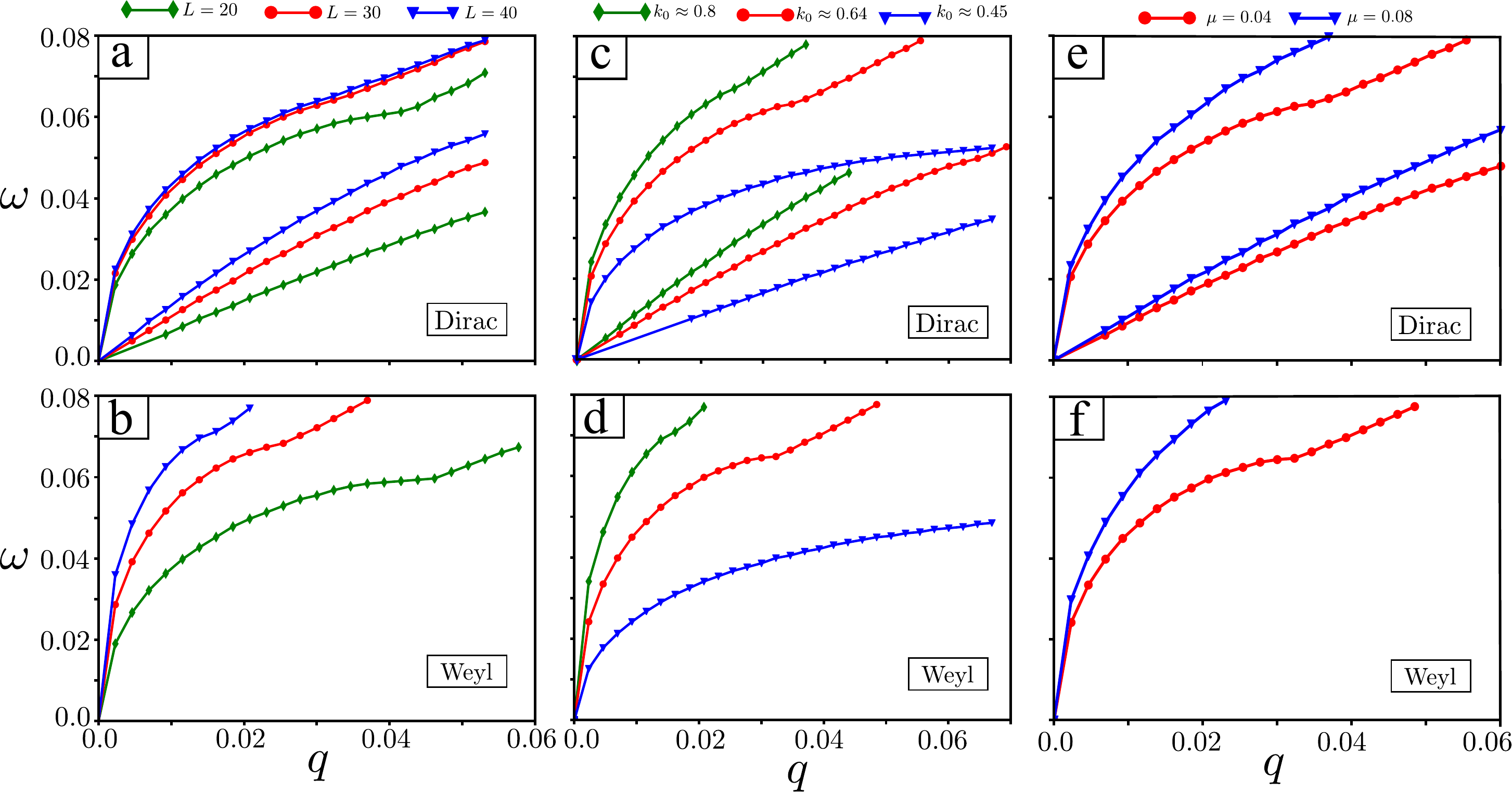}
	\caption{(a), (b): Plasmon dispersions for few values of the thickness $L$; (c), (d): Plasmon dispersions for few values of the distance between the Weyl nodes, given by  2$k_0$ and  (e), (f): Plasmon dispersions for two values of the chemical potential $mu$. All other parameters, in each cases, are the same as in Fig.~2 of the main text.}\label{fig:variation}
\end{figure*}

\subsection*{D. Further properties of the Plasmon modes}
\subsubsection*{Variation with $\theta=\cos^{-1}q_x/q$}
The results from the simple heuristic model, Eq.~(\ref{eq:pld}) and Eq.~(\ref{eq:plw}), predict that the plasmon disperses more slowly with increasing $\theta$. To test this we numerically computed the plasmon dispersions for a range of $\theta$.  The results are plotted in Figs.~\ref{fig:wtheta}(a) and (b), essentially verifying this expectation. Furthermore, for a given value of $\mb{q}$, the quantity -Im(Tr($\chi(\mb{q},\omega)$)) captures the collective mode density of states. 
Figs.~\ref{fig:wtheta} (c) and (d) illustrate how this density of states behaves in presence of the sharp plasmon mode. Note that widths of the peaks at the plasmon mode frequencies are due to an infinitesimal imaginary part added to the frequency for the calculation of the response function.

\subsubsection*{Localization of the plasmon mode}
When the condition for a plasmon mode is met, one of the eigenvalues of the matrix $(I - \Delta z B(\mb{q},\omega))$ vanishes. At this value of $\mb{q}$ and $\omega$, $\chi$ is fully dictated by the eigenvector corresponding to the vanishing eigenvalue. This can be understood in the following way: the inverse of the matrix $(I - \Delta z B(\mb{q},\omega))$ can be written in the basis of eigenvectors in the form
\begin{align}
(I - \Delta z B(\mb{q},\omega))^{-1} = \sum_{\lambda}\frac{|\lambda\rangle\langle\lambda|}{\lambda}.
\end{align}
Thus when one of the eigenvalue, say $\lambda_0$, approaches 0, the sum over eigenvalues is dominated by this contribution, so that
\begin{align}
\chi \approx \frac{|\lambda_0\rangle\langle \lambda_0|}{\lambda_0} \chi_0.
\end{align}
The real space density associated with the relevant eigenvector as a function of $z$, $\psi(z)=\langle z|\eta\rangle$, indicates whether the density oscillations of the mode are symmetric or antisymmetric with respect to  $z$, or somewhere between these behaviors.   We plot some representative densities as a function of $z$ in Fig.~\ref{fig:wtheta}(e)-(g). For small $q$, the modes have relatively equal support on the two surfaces as well as substantial support in the bulk. For larger $q$, the modes are more localized near one of the surfaces.

\subsubsection*{Variation with other parameters of the model}
In Fig.~\ref{fig:variation} we show how the plasmon dispersions vary as a function of the relevant scales of the problem, such as the thickness of the film ($L$), the length of the Fermi arc (given by 2$k_0$) in the momentum space, and the chemical potential $\mu$. In all these calculations we assumed the chemical potential is below the $n=2$ band, so that there is a gap in the particle-hole continuum for small $q$. With increasing thickness $L$, if the chemical potential is still below the $n=2$ band, we observe that the dispersions of the plasmon modes become steeper. This can be attributed to more localized surface states with larger thickness. In case of the WSM, when $L\rightarrow\infty$, one expects to recover the results for a single Fermi arc, where the plasmon mode is gapped, as predicted in Eq.~(\ref{eq:singleWFA}).

An increase in the distance between the Weyl nodes (given by $2k_0 = 2\cos^{-1}m$) while keeping other parameters the same increases the localization of the Fermi arc states on the surfaces, as well as increases the surface density of states.  This results in steeper dispersions for the plasmon modes, which is also evident from Eq.~(\ref{eq:pld}) and  Eq.~(\ref{eq:plw}).

On the other hand, increasing chemical potential $\mu$, keeping other parameters the same, increases the size of the Fermi-surface (see Fig.~1 of the main text), as long as the chemical potential remains smaller than the next band. This allows a longer FA. As it is clear from Eq.~(\ref{eq:pld}), a longer FA is predicted to result in steeped plasmon dispersion (such length enters through the parameter $\beta$), which is also numerically obtained as shown in Fig.~\ref{fig:variation} (e), (f).

When the chemical potential exceeds the bottom of the $n=2$ band, bulk states begin to screen the surface modes more effectively, which further weakens the dispersions of the surface plasmon modes
that are the focus of our study.  With increasing $\mu$ one expects the surface plasmons to ultimately merge with bulk plasmons. We leave a full characterization of this evolution from surface to bulk plasmons for future research.


\begin{thebibliography}{99}
	\bibitem{WSMreview}
	For reviews see: N.~P.~Armitage, E.~J.~Mele, and A.~Vishwanath, Rev. Mod. Phys. \textbf{90}, 015001 (2018); Nature S. Jia, S.-Y. Xu, and M. Z. Hasan, Nature Materials \textbf{15}, 1140–1144 (2016); S. Rao, Journal of the Indian Institute of Science, \textbf{96}, 2 (2016).
	
	\bibitem{taas}
	B.~Q.~Lv, H.~M.~Weng, B.~B.~Fu, X.~P.~Wang, H.~Miao, J.~Ma, P.~Richard, X.~C.~Huang, L.~X.~Zhao, G.~F.~Chen, Z.~Fang, X.~Dai, T.~Qian, and H.~Ding, Phys. Rev. X \textbf{5}, 031013 (2015).

	\bibitem{nbas}
	Zhang, C., Ni, Z., Zhang, J. \textit{et al.}, Nat. Mater. 18, 482–488 (2019).

	\bibitem{CoSnS1}
	Guowei Li \textit{et al.}, Science Advances Vol. 5, no. 8 (2019).

	\bibitem{CoSnS2}
	M. Tanaka \textit{et al.}, Nano Lett. 2020, 20, 10, 7476–7481.


	\bibitem{cdas0}
	I. Crassee, R. Sankar, W.-L. Lee, A. Akrap, and M. Orlita, Phys. Rev. Materials \textbf{2}, 120302 (2018).
	

	\bibitem{cdas1}
	Zhang, C., Narayan, A., Lu, S. \textit{et al.}, Nature Communications \textbf{8}, 1272 (2017).

	\bibitem{cdas2}
	Timo Schumann, Luca Galletti, David A. Kealhofer, Honggyu Kim, Manik Goyal, and Susanne Stemmer, Phys. Rev. Lett. \textbf{120}, 016801 (2018).

	\bibitem{cdas3}
	Manik Goyal, Luca Galletti, Salva Salmani-Rezaie, Timo Schumann, David A. Kealhofer, and Susanne Stemmer, APL Mater. \textbf{6}, 026105 (2018).

	\bibitem{weylpl1}
	Johannes Hofmann and S. Das Sarma, Phys. Rev. B 91, 241108 (2015).

	\bibitem{weylpl2}
	Krishanu Sadhukhan, Antonio Politano, and Amit Agarwal, Phys. Rev. Lett. 124, 046803 (2020).

	\bibitem{weylplsurf1}
	Johannes Hofmann and Sankar Das Sarma, Phys. Rev. B \textbf{93}, 241402(R) (2016).

	\bibitem{weylplsurf2}
	Gian Marcello Andolina, Francesco M. D. Pellegrino, Frank H. L. Koppens, and Marco Polini, Phys. Rev. B \textbf{97}, 125431 (2018).

	\bibitem{weylplsurf3}
	Gennaro Chiarello, Johannes Hofmann, Zhilin Li, Vito Fabio, Liwei Guo, Xiaolong Chen, Sankar Das Sarma, and Antonio Politano, Phys. Rev. B 99, 121401 (2019).

	\bibitem{weylplsurf4}
	Kota Tsuchikawa, Satoru Konabe, Takahiro Yamamoto, and Shiro Kawabata, Phys. Rev. B 102, 035443 (2020).

	\bibitem{weylplsurf5}
Justin C. W. Song and Mark S. Rudner, Phys. Rev. \textbf{B} 96, 205443 (2017).

\bibitem{weylplsurf6}
 Željana Bonačić Lošić 2018 J. Phys.: Condens. Matter \textbf{30}, 365003 (2018).

\bibitem{weylplsurf7}
Tomohiro Tamaya, Takeo Kato, Satoru Konabe, and Shiro Kawabata, J. Phys.: Condens. Matter \textbf{31}, 305001 (2019).

\bibitem{weylplsurf8}
S. Oskoui Abdol, A. Soltani Vala, and B. Abdollahipour,  J. Phys.: Condens. Matter \textbf{31} (2019).

\bibitem{rkky}
Sonu Verma, Debasmita Giri, H.A. Fertig, and Arijit Kundu, Phys. Rev. B \textbf{101}, 085419 (2020).

	\bibitem{twolr1}
	H. Gutfreund and Y. Unna, J. Phys. Chem. Solids, (1973).

	\bibitem{twolr2}
	S. Das Sarma and A. Madhukar, Phys. Rev. B {\bf 23}, 805 (1981).

	\bibitem{twolr3}
	L. Liu, L. Swierkowski, D. Neilson, and J. Szymanski, Phys. Rev.B \textbf{53}, 7923 (1996).

	\bibitem{twolr4}
	E. H. Hwang and S. Das Sarma, Phys. Rev. B \textbf{80}, 205405 (2009).

	\bibitem{twolr5}
	T. Stauber, G. Gómez-Santos, and L. Brey, ACS Photonics \textbf{4}, 12, 2978–2988 (2017).

	\bibitem{twolr6}
	Z. Jalali-Mola and S. A. Jafari, Phys. Rev. B (2018).

	\bibitem{twolr7}
	Rajdeep Sensarma, E. H. Hwang, and S. Das Sarma,
	 Phys. Rev. B \textbf{82}, 195428 (2010).

\bibitem{Hosur_2013}
Pavan Hosur and Xiaoliang Qi, Comptes Rendus Physique {\bf 14}, 857 (2013).

\bibitem{ShuoWang_2017}
Shuo Wang, Ben-Chuan Lin, An-Qi Wang, Da-Peng Yu and Zhi-Min Liao,
Advances in Physics: X {\bf 2}, 518 (2017).

\bibitem{Wang_2018} H. Wang and J. Wang, Chinese Phys. B {\bf 27}, 107402 (2018).

    \bibitem{Chen_2012} J. Chen et al., Nature {\bf 487}, 77 (2012).

    \bibitem{Fei_2012} Z. Fei et al., Nature {\bf 487}, 82 (2012).


\bibitem{alpha}
For a generic material, $\alpha$ is given by $\alpha \approx \frac{c}{v_f}\times\frac{1}{137}$.

\bibitem{wvector}
Essentially, at resonance (i.e, when det$[I - \Delta z B(\mb{q},\omega)]=0$ is satisfied), the eigenvector of the matrix $B(\mb{q})$ is also the eigenvector of the response matrix.




\bibitem{landau}
J. B. Khurgin, G. Sun, in Quantum  Plasmonics, S. Bozhevolnyi,L. Martin-Moreno, F. Garcia-Vidal, Eds. (Springer, 2016), ch. 13, pp. 303–322.

\bibitem{Tanaka_2020} Shunsuke Tanaka, Tatsuya Yoshida, Kazuya Watanabe, Yoshiyasu Matsumoto, Tomokazu Yasuike, Marin Petrović, and Marko Kralj, Phys. Rev. Lett. {\bf 125}, 126802 (2020).




	
\end{thebibliography}
\end{document}